\documentclass[aps,prd,superscriptaddress,amsmath,twocolumn,floatfix,showpacs]{revtex4-1}
\newcommand{\beq}{\begin{equation}}
\newcommand{\eeq}{\end{equation}}
\newcommand{\beqn}{\begin{eqnarray}}
\newcommand{\eeqn}{\end{eqnarray}}
\newcommand{\aap}{Astron.\ and\ Astrophys.\ }

\newcommand{\apjs}{Astrophys.\ J.\ Supp.}
\newcommand{\mnras}{Month. Not. Roy. Astron. Soc.}

\newcommand{\llabel}[1]{\label{#1}}              
\newcommand{\labeq}[2]{ \begin{equation} \llabel{#1}
{#2}
\end{equation}}
\usepackage{graphicx}
\usepackage{epsf}
\usepackage{graphics,epsfig,placeins,subfigure,wrapfig,amssymb,pifont}
\begin{document}
\title{Relativistic MHD in dynamical spacetimes:\\
Improved EM gauge condition for AMR grids}

\author{Zachariah B. Etienne}
\email{zetienne@illinois.edu}
\author{Vasileios Paschalidis}
\author{Yuk Tung Liu}
\author{Stuart~L.~Shapiro}
\altaffiliation{Also at Department of Astronomy and NCSA, University of
  Illinois at Urbana-Champaign, Urbana, IL 61801}
\affiliation{Department of Physics, University of Illinois at
  Urbana-Champaign, Urbana, IL 61801}

\begin{abstract}

We recently developed a new general relativistic magnetohydrodynamic code
with adaptive mesh refinement that evolves the
electromagnetic (EM) vector potential $A_i$ instead of the magnetic fields
directly.  Evolving $A_i$ enables one to use any interpolation scheme on
refinement level boundaries and still guarantee that the magnetic field remains
divergenceless. As in classical EM, a gauge choice must 
be made when evolving $A_i$, and we chose a straightforward ``algebraic'' 
gauge condition to simplify the $A_i$ evolution equation. However, 
magnetized black hole-neutron star 
(BHNS) simulations in this gauge exhibit
unphysical behavior, including the spurious appearance of
strong magnetic fields on refinement level boundaries.  This spurious
behavior is exacerbated when matter crosses refinement boundaries
during tidal disruption of the NS.  Applying
Kreiss-Oliger dissipation to the evolution of the magnetic vector
potential $A_i$ slightly weakens this spurious magnetic effect, but
with undesired consequences.  We
demonstrate via an eigenvalue analysis and a numerical study that
zero-speed modes in the algebraic gauge, coupled with the
frequency filtering that occurs on refinement level boundaries, are
responsible for the creation of spurious magnetic fields.  
We show that the EM Lorenz gauge exhibits no
zero-speed modes, and as a consequence, spurious magnetic effects
are quickly propagated away, allowing for long-term,
stable magnetized BHNS evolutions. Our study demonstrates 
how the EM gauge degree of freedom can be chosen 
to one's advantage, and that for magnetized BHNS simulations 
the Lorenz gauge constitutes a major improvement over the algebraic gauge.

\end{abstract}


\maketitle

\section{Introduction}

Magnetized fluids in dynamical, strongly curved spacetimes
play a central role in many systems of current interest
in relativistic astrophysics. 
The presence of magnetic fields may destroy 
differential rotation in nascent neutron stars, 
form jets and influence disk dynamics around black holes, affect
collapse of massive rotating stars, etc. Many of
these systems are promising sources of gravitational radiation
for detection by laser interferometers such as LIGO,
VIRGO, TAMA, GEO, LCGT, ET, LISA and DECIGO. Some may also emit electromagnetic
radiation as gamma-ray bursts, or emission from magnetized
disks around black holes in active galactic nuclei and quasars. Accurate,
self-consistent modeling of these systems requires computational schemes 
capable of simultaneously accounting for magnetic fields, 
relativistic magnetohydrodynamics (MHD), radiation transport and 
relativistic gravitation.

Over the past several years, we have developed a numerical scheme 
in 3+1 dimensions capable of solving the coupled system of 
Einstein's field equations, relativistic MHD and 
Maxwell's equations~\cite{DSY04,ELS2010}. Our approach is based on
the BSSN (Baumgarte-Shapiro-Shibata-Nakamura) formalism
to evolve the metric \cite{SN,BS}, a high-resolution,
shock-capturing (HRSC) scheme to handle the fluids,
and a constrained-transport (CT) scheme to treat the magnetic
induction equation~\cite{toth}. This GRMHD code 
has been subjected to a rigorous suite of numerical tests to 
check and calibrate its accuracy \cite{DSY04,ELS2010}. 
The code has been applied to
explore a number of important dynamical scenarios in
relativistic astrophysics, including the collapse of magnetized,
differentially rotating hypermassive neutron stars to black holes 
\cite{dlsss06a,dlsss06b,SSL08}, the collapse of rotating stellar cores
to neutron stars \cite{ShiUIUC}, the collapse of rotating, supermassive
stars and massive Population III stars to black holes \cite{lss07},
magnetized binary neutron star merger \cite{lset08}, binary black
hole-neutron stars (BHNSs) \cite{eflstb08,elsb09}, binary white 
dwarf-neutron stars \cite{PELS11,PLES11},
and the merger of binary black holes in 
gaseous clouds \cite{FLS10} and disks \cite{FLS11}.

Many problems in relativistic astrophysics require numerical
simulations covering a large range of length scales.
For example, to follow the final merger of a compact
binary system with a total mass $M$, length scales of $M/30$ or less
must be resolved to treat the strong-field,
near-zone regions reliably. On the other hand, accurate
gravitational wave calculations at length scales $\sim M$ must
be performed far in the weak-field wave-zone at radius
$r \gtrsim 100M$. Adaptive mesh refinement (AMR) allows for 
sufficient resolution to be
supplied to areas of the computational domain as needed,
enabling us to resolve both strong- and weak-field domains
efficiently.

One of the most subtle issues in evolving the MHD
equations is the preservation of the divergenceless constraint
($\mathbf{\nabla\cdot B} = 0$). When evolving
the induction equation, numerical truncation error leads
to violations of this constraint, resulting
in unphysical plasma transport orthogonal to the magnetic
field, as well as violations of energy and momentum
conservation (see e.g., \cite{toth,BB80,BS99}). In simulations
using a uniformly-spaced grid, constrained-transport 
schemes (see e.g., \cite{eh88,toth}) are commonly used to maintain
the divergenceless constraint. In simulations using AMR
grids, both CT schemes and the hyperbolic
divergence-cleaning (HDC) method \cite{DKKMSW02,AHLS06} have been
used. 
However, we find that in the presence of black holes, otherwise
stable GRMHD evolution schemes often become unstable.

One of the most commonly adopted methods for evolving
black holes is the moving puncture technique \cite{RIT1,God1}, in
which the puncture gauge conditions ensure that the spacetime
singularity in the black hole interior never appears.
However, a {\it coordinate} singularity is present in the
computational domain near which accurate numerical evolution
is difficult to achieve. It has been demonstrated
that when using the moving puncture gauge, while gauge modes can 
leak out of the BH horizon \cite{Brown09},
any physical or constraint violating
data in the black hole interior will not propagate out of the horizon
\cite{EFLSB,Turducken,Turducken2}. 
We find that this property is
preserved in the presence of hydrodynamic matter, but can be challenging to
maintain when the matter is magnetized. 
Though we have found that the HDC scheme works well with the moving
puncture technique in 
the absence of black holes,
when BHs are present and excision is not applied, 
we find that, even in the Cowling
approximation, in which the metric is fixed, inaccurate data 
in the black hole interior can propagate out of the horizon and
contaminate the solutions. 

For this reason, in developing a moving puncture-compatible 
algorithm for maintaining
$\mathbf{\nabla\cdot B} = 0$, we focused on CT schemes. 
This was the approach adopted in our earlier, unigrid GRMHD
code~\cite{DSY04}. In AMR simulations, the conventional CT
scheme may be used where individual refinement
levels do not overlap. However, maintaining the divergenceless constraint
at refinement level boundaries requires that special interpolations
be performed during prolongation and restriction.
Such prolongation/restriction operators have been devised
\cite{Balsara01,Balsara09,McNally11}, but these operators must be
fine-tuned to the particular AMR implementation, and some have only been
tested for Newtonian MHD.

In \cite{ELS2010} we developed an alternative, AMR-compatible CT
scheme. The scheme is based on the CT
method described in~\cite{zbl03}, which recasts
the magnetic induction equation as an evolution
equation for the magnetic vector potential.  Divergence-free magnetic
fields are guaranteed as they are computed from
the curl of the vector potential. The evolution of the vector potential
is carried out in the same HRSC framework as other
hydrodynamic variables. This scheme is numerically
equivalent to the commonly-used, staggered-$B$ CT
scheme of~\cite{BS99} for uniform resolution simulations, but is readily
generalized to an AMR grid. Unlike the
magnetic field, the vector potential is not constrained,
so any interpolation scheme may be used during prolongation
and restriction, thus enabling its use with any AMR algorithm.
In addition, since the vector potential does not uniquely
determine the magnetic field, there exists an extra gauge degree
of freedom that can be used to one's advantage. A
straightforward ``algebraic'' gauge condition was adopted
in our original formulation, to simplify evolution equation
for the vector potential (see also \cite{zbl03,grb11}).

In \cite{ELS2010} we performed several tests on our new AMR-compatible 
CT scheme. We found that our scheme works well even in black-hole spacetimes. 
Inaccurate data generated in the black hole interior stayed inside the horizon. 
Hence our scheme is compatible with the moving puncture technique.
However, when this algebraic gauge condition is used for magnetized binary 
black hole--neutron star (BHNS) simulations, stable, long-term
evolutions are not possible.

In this paper, we exploit another EM gauge condition, 
the Lorenz gauge~\footnote{Note that the EM Lorenz gauge condition is named after Ludvig Lorenz
and not after Hendrik Lorentz known for the Lorentz transformations (see also note in p. 294 in 
\cite{Jackson3rd}).}, which allows for stable, 
long-term GRMHD evolutions of binary BHNSs. 
Performing eigenvalue analyses, we find
that static EM modes are present when adopting the algebraic gauge, 
whereas all EM modes are propagating when the Lorenz gauge is adopted. 
As a result, when the algebraic gauge is used, its 
zero-speed modes are inevitably excited 
and remain on the grid. Further, when these modes are interpolated 
at refinement boundaries, inaccuracies in interpolation may lead to
the appearance of spurious magnetic fields that
contaminate the solution and lead to a build-up of
numerical error.  
Given that the Lorenz gauge possesses no static modes, any such spurious 
modes should be propagated away to the boundary and 
leave the grid. Thus, we expect that the solutions will 
be improved when using the Lorenz gauge.

Using magnetized BHNS simulations
we perform a comparison between the Lorenz gauge and the algebraic gauge
with and without Kreiss-Oliger dissipation (KOD) applied 
on the vector potential evolution.
We confirm the expected behavior. The 
BHNS system that forms the basis of our 
comparison is an irrotational binary in an initially circular orbit, 
with a BH:NS mass ratio of 3:1 and the NS 
initially seeded with a small poloidal magnetic field.
Using these simulations we demonstrate the superiority of 
the Lorenz gauge, and confirm that in the algebraic
gauge static modes are present and lead to spurious 
B-fields which remain on the grid. This spurious effect 
contaminates the solution and is amplified in time, leading the simulations 
to break down shortly after NS tidal disruption. In contrast 
to the algebraic gauge, 
static EM modes are not present in the Lorenz gauge, and any spurious fields
generated at refinement level boundaries 
are quickly propagated away. Overall we find that the 
Lorenz gauge provides a major improvement  
for magnetized BHNS simulations. In~\cite{elpb11}, we employ 
the Lorenz gauge to perform 
a detailed study of the effects of magnetic fields 
in the evolution of binary BHNSs.

The remaining sections are organized as follows.
In Sec.~\ref{sec:basic_eqns} 
we outline the basic GRMHD equations and provide an 
overview of EM gauge conditions and evolution equations.
In Sec.~\ref{sec:analysis} we perform an eigenvalue analysis of the 
EM evolution equations. In Sec.~\ref{sec:numerical},  we briefly 
describe the specific implementation of our GRMHD scheme. 
In Sec.~\ref{sec:results}, we review the differences between 
our magnetized BHNS simulations in the algebraic and Lorenz gauges.
We summarize our findings in Sec.~\ref{sec:discussion}.

\section{Basic Equations}
\label{sec:basic_eqns}

The formulation and numerical scheme for our simulations is basically
the same as in our previous work~\cite{eflstb08,elsb09,ELS2010}, to which
the reader may refer for details.  Here we introduce our notation,
summarize our method, and point out the latest changes to our
numerical technique.  Geometrized units ($G = c = 1$) are adopted,
except where stated explicitly.  Greek indices denote all four
spacetime dimensions (0, 1, 2, and 3), and Latin indices imply spatial
parts only (1, 2, and 3).

We use the 3+1 spacetime decomposition, in which  
the metric has the following form:
\beq
  ds^2 = -\alpha^2 dt^2
+ \gamma_{ij} (dx^i + \beta^i dt) (dx^j + \beta^j dt) \ ,
\eeq
where $\alpha$ is the lapse function, $\beta^i$ the shift vector,
and $\gamma_{ij}$ the three-metric on spacelike hypersurfaces of constant 
time $t$. We then decompose all evolution equations in 3+1 form 
(see e.g. \cite{BSBook} for a detailed discussion and references).

\subsection{Gravitational fields}

We adopt the Baumgarte-Shapiro-Shibata-Nakamura (BSSN) 
formalism~\cite{SN,BS} in which 
the metric evolution variables are the conformal exponent $\phi
\equiv \ln (\gamma)/12$, the conformal 3-metric $\tilde
\gamma_{ij}=e^{-4\phi}\gamma_{ij}$, three auxiliary functions
$\tilde{\Gamma}^i \equiv -\tilde \gamma^{ij}{}_{,j}$, the trace $K$ of
the extrinsic curvature $K_{ij}$, and the trace-free part of 
the conformal extrinsic curvature $\tilde A_{ij} 
\equiv e^{-4\phi}(K_{ij}-\gamma_{ij} K/3)$.
Here, $\gamma={\rm det}(\gamma_{ij})$. The full spacetime metric $g_{\mu \nu}$
is related to the three-metric $\gamma_{\mu \nu}$ by $\gamma_{\mu \nu}
= g_{\mu \nu} + n_{\mu} n_{\nu}$, where the future-directed, timelike
unit vector $n^{\mu}$ normal to the time slice can be written in terms
of the lapse $\alpha$ and shift $\beta^i$ as $n^{\mu} = \alpha^{-1}
(1,-\beta^i)$. The evolution equations of these BSSN variables are 
given by Eqs.~(9)--(13) of~\cite{eflstb08}. 
We adopt the standard puncture gauge conditions: an advective
``1+log'' slicing condition for the lapse and a 
``Gamma-freezing'' condition for the shift~\cite{GodGauge}. The 
evolution equations for $\alpha$ and $\beta^i$ are given by 
Eqs.~(2)--(4) of~\cite{elsb09}, with the 
$\eta$ parameter set to $\approx 2.2/M$ in all 
BHNS simulations presented here, where $M$ is the ADM mass of the binary. 
We add a sixth-order KOD term of the form 
$\epsilon/64 (\Delta x^5 \partial_x^{(6)}+
\Delta y^5 \partial_y^{(6)}+\Delta z^5 \partial_z^{(6)})f$  
to reduce high-frequency numerical noise associated with AMR refinement
interfaces. Here $f$ stands for all evolved BSSN, lapse and shift 
variables. We choose the strength parameter as $\epsilon=0.2$. 
Note that this KOD term converges to zero at fifth-order in the grid 
spacing so that it does not alter the convergence properties 
of our BSSN scheme.

\subsection{Magnetohydrodynamic fields}

The fundamental MHD variables include the rest-mass density 
$\rho_0$, specific internal energy $\epsilon$, pressure $P$, 
four-velocity $u^{\mu}$, and magnetic field $B^\mu=n_{\nu} F^{* \nu \mu}$. 
Here $F^{* \mu \nu}$ is the dual of the Faraday tensor $F^{\mu \nu}$. 
Note that $B^\mu$ is purely spatial ($B^0 =-n_\mu B^\mu/\alpha=0$). 
We adopt a $\Gamma$-law equation of state (EOS)
$P=(\Gamma-1)\rho_0 \epsilon$ with $\Gamma=2$, which reduces to 
an $n=1$ polytropic law for the initial (cold) neutron star matter. 
The stress-energy tensor is given by 
\beq
  T_{\mu \nu} = (\rho_0 h +b^2) u_\mu u_\mu 
+ \left(P +\frac{b^2}{2}\right) g_{\mu \nu} - b_\mu b_\nu \ ,
\eeq
where $h=1+\epsilon+P/\rho_0$ is the specific enthalpy and 
\beq
  b_\mu = -\frac{P_{\mu \nu} B^\nu}{\sqrt{4\pi}\, n_{\nu} u^\nu}
\label{eq:bmu}
\eeq
is the magnetic field measured in fluid's comoving frame, modulo 
a factor of $1/\sqrt{4\pi}$. Here $P_{\mu \nu}=g_{\mu \nu} + u_\mu u_\nu$ 
and $b^2=b^\mu b_\mu$. In the ideal MHD limit, in which the plasma is 
assumed to have perfect conductivity, the Faraday tensor can be written 
as $F^{\mu \nu} = \sqrt{4\pi}\, u_{\gamma} \epsilon^{\gamma \mu \nu \delta} b_\delta$.

In the standard numerical implementation of the MHD
equations using a conservative scheme, it is useful to introduce the 
``conservative'' variables 
$\rho_*$, $\tilde{S}_i$, $\tilde{\tau}$ and $\tilde{B}^i$. They are 
defined as 
\beqn
&&\rho_* \equiv - \sqrt{\gamma}\, \rho_0 n_{\mu} u^{\mu} \ ,
\label{eq:rhos} \\
&& \tilde{S}_i \equiv -  \sqrt{\gamma}\, T_{\mu \nu}n^{\mu} \gamma^{\nu}_{~i}
\ , \\
&& \tilde{\tau} \equiv  \sqrt{\gamma}\, T_{\mu \nu}n^{\mu} n^{\nu} - \rho_* \ , 
\label{eq:S0} \\
&& \tilde{B}^i \equiv \sqrt{\gamma}\, B^i .
\label{eq:Btilde}
\eeqn
The evolution equations for $\rho_*$, $\tilde{S}_i$ and $\tilde{\tau}$ can 
be derived from the conservation of baryon number $\nabla_\mu (\rho_* u^\mu)=0$ 
and conservation of energy-momentum $\nabla_\mu T^{\mu \nu}=0$, giving rise to 
Eqs.~(27)--(30) in~\cite{ELS2010}. 

\subsection{Electromagnetic fields}

In the ideal MHD limit, the Maxwell equation 
$\nabla_\nu F^{* \mu \nu}=0$ yields the magnetic 
constraint $\partial_j \tilde{B}^j=0$ 
and the induction equation
 $\partial_t \tilde{B}^i + \partial_j (v^j \tilde{B}^i 
- v^i \tilde{B}^j)=0$. As shown in~\cite{ELS2010} and~\cite{bs03},
these equations can be rewritten 
by introducing the electromagnetic 4-vector potential 
${\cal A}_\mu = \Phi n_\mu+ A_\mu$, with $n^\mu A_\mu=0$. 
The magnetic constraint and induction equations become 
\beqn
  B^i &=& \epsilon^{ijk} \partial_j A_k \ , \label{eq:BfromA} \\ 
  \partial_t A_i &=& \epsilon_{ijk} v^j B^k 
- \partial_i (\alpha \Phi -\beta^j A_j) \ 
\label{eq:Aevol}
\eeqn
where $\epsilon^{ijk}=n_\mu \epsilon^{\mu ijk}$ is the 3-dimensional 
Levi-Civita tensor. 
The dynamical variable in this scheme is the 
vector potential. The B-field in Eq. \eqref{eq:Aevol} and in the 
MHD equations is essentially replaced according to Eq. \eqref{eq:BfromA}.

The EM evolution via the vector potential introduces 
an additional gauge degree of freedom. Therefore, an EM gauge choice must
be made to close the system of equations. 

\subsubsection{Algebraic gauge}

In~\cite{zbl03,ELS2010}, the algebraic EM gauge condition 
\labeq{algebraic}{
\Phi=\frac{1}{\alpha}\beta^j A_j=-n^j A_j
}
is adopted because it simplifies the evolution of $A_i$ to
\labeq{}{
  \partial_t A_i = \epsilon_{ijk} v^j B^k.
}

\subsubsection{Lorenz gauge}

In this paper, we find that simulations of magnetized BHNSs using AMR and the algebraic gauge 
suffer from severe numerical errors and eventually break down due to
a build-up of errors.

Improved, stable, long-term 
BHNS evolutions may be achieved by imposing the Lorenz gauge condition
\labeq{Lorenz_standard}{
\nabla_\mu {\cal A}^\mu = 0,
}
which yields the evolution equation 
\beq
  \partial_t (\sqrt{\gamma}\, \Phi) + \partial_j (\alpha \sqrt{\gamma}\, A^j 
- \sqrt{\gamma}\, \beta^j \Phi) = 0 
\label{eq:Phievol}
\eeq
for $\Phi$.

Our numerical implementation of
Eqs.~(\ref{eq:BfromA}) and (\ref{eq:Aevol}) guarantees numerically
identical $B^i$ regardless of EM gauge
in simulations with a uniform-resolution grid.  However,
interpolations performed on $A_i$ at refinement boundaries on AMR
grids will modify $A_i$, resulting in different $B^i$ on and near these
boundaries, ultimately leading to different stability properties.

\section{Electromagnetic gauge analysis}
\label{sec:analysis}

The differences in the stability properties of the 
gauges considered here can be explained
by the fact that the algebraic gauge possesses a 
static mode, whereas all modes in 
the Lorenz gauge are propagating. Here we perform an 
eigenvalue analysis of the EM field evolution equations with 
the algebraic and Lorenz gauges and demonstrate 
explicitly the existence of static modes in the former gauge.

\subsection{Evolution Equations}

Substituting Eq.~\eqref{eq:BfromA} into Eq.~\eqref{eq:Aevol} we find that 
the evolution of the vector potential is given by
\labeq{tAb}{
\begin{split}
\partial_t A_i = &\ -v^m\partial_mA_i + (v^m+\beta^m) \partial_i A_m 
                -\alpha \partial_i \Phi \\
                &\  - \Phi \partial_i \alpha + A_m \partial_i \beta^m .
\end{split}
}

Writing Eq.~\eqref{eq:Phievol} as an evolution equation for $\Phi$ we find
\begin{eqnarray}
\partial_t \Phi & = & \beta^m\partial_m\Phi - 
                     \alpha\partial_mA^m \nonumber 
                    +\gamma^{-1/2}\Phi\partial_m( \sqrt{\gamma}\beta^m) \\
                    & &  - \gamma^{-1/2}A^m\partial_m (\alpha\sqrt{\gamma})
                    -\Phi\partial_t \ln(\sqrt{\gamma}). \nonumber \\     
\end{eqnarray}
 
In the limit of short-wavelength, small-amplitude
perturbations about a background solution, the EM evolution 
equations decouple from the MHD and BSSN equations
and can be analyzed separately. While the entire coupled system of 
the BSSN, MHD and Maxwell's equations is quasi-linear, 
the $A_i, \Phi$ partial differential 
equations alone are linear, as the coefficients 
that multiply the spatial derivatives 
of $A_i$ and $\Phi$ are not functions of $A_i$ and $\Phi$. 
For our purpose we can apply the hyperbolicity theorems 
for linear systems of equations \cite{GKOpdeBook,KLpdeBook}, 
thus we need only consider the principal 
parts of the evolution equations and the localization principle.

\subsection{Algebraic gauge}

In this gauge only the vector potential evolves 
dynamically and $\Phi$ is entirely determined by $A_i$ and
the BSSN gauge variables. The principal part of the 
$A_i$ evolution equation is given by
\labeq{}{
\partial_t A_i \simeq -v^m\partial_m A_i + v^m\partial_i A_m.
}
Here $\simeq$ denotes ``the principal part equals''.
Replacing $\partial_i$ by $k_i$, where $k_i$ is the spatial component 
of the wave vector, normalized according to 
$\gamma^{ij}k_i k_j=1$, we obtain
\labeq{}{
\partial_t A_i \simeq -v^m k_m A_i + v^mk_i A_m \equiv F_i
}
We can now construct the characteristic matrix of the 
system: $M_{ij} = \partial F_i/\partial A_j$. For the system 
to be strongly hyperbolic the eigenvalues of this matrix 
must be real and the matrix 
must possess a complete set of eigenvectors
\footnote{Strictly speaking, 
for the quasi-linear, partial differential equations considered here,
the existence of real eigenvalues and 
a complete set of eigenvectors are only necessary conditions 
for strong hyperbolicity (see \cite{GKOpdeBook,KLpdeBook} for 
complete definitions of strong hyperbolicity). However, here 
we are not interested in rigorously 
proving strong hyperbolicity. We are only interested 
in identifying the eigenvalues of the characteristic matrix and, in particular,
whether static modes exist or not.}. A straightforward calculation shows that 
the set of eigenvectors of this system is complete and the eigenvalues (or 
characteristic speeds) are
\labeq{algmodes}{
\lambda_1 = 0,\ \lambda_{2,3} =  v^m k_m.
}
Hence the system is strongly hyperbolic. 
However, the system possesses a static gauge mode, i.e., 
a mode that does not propagate. In unigrid simulations, this is not an
issue since our numerical implementation of the $A_i$ evolution
guarantees $B^i$ (obtained via Eq.~(\ref{eq:BfromA})) to be equivalent
to the standard staggered-$B$ CT scheme \cite{zbl03}, which evolves $B^i$ directly and thus
does not suffer from potential EM gauge mode problems.
In AMR simulations however, $A_i$ is interpolated between refinement
levels. The static modes will remain in the computational domain where
they were generated, allowing 
interpolation errors between refinement 
levels to give rise to spurious $B^i$ at refinement level boundaries, 
which subsequently propagate to the interior of the refinement boxes 
and affect the dynamics of system. 
In this way, numerical errors can build up and lead to unstable behavior.
On a related note, we point out that the existence of zero modes has also been considered
as an explanation for the different stability properties of different 
3+1 formulations of GR \cite{AABSS2000,SY2002,PV2008,PHK2008}. 
However, the existence of zero speed modes is not always associated
with unstable behavior. For example the BSSN formulation gives 
rise to stable numerical integrations even though 
it possesses a zero-speed mode.

\subsection{Lorenz gauge}

In this gauge both the vector and scalar potentials evolve. 
The principal part of the EM evolution equations in the Lorenz gauge is
\labeq{}{\begin{split}
\partial_t A_i \simeq &\-v^m\partial_m A_i + (v^m+\beta^m) \partial_i A_m 
                      - \alpha \partial_i \Phi, \\
                      \partial_t \Phi \simeq &\ \beta^m \partial_m \Phi
                      - \alpha \gamma^{mn}\partial_m A_n.
\end{split}
}
Using the $\partial_i \rightarrow k_i$ rule and 
calculating the eigensystem of the characteristic 
matrix we find that the matrix has a complete 
set of eigenvectors and its eigenvalues are
\labeq{lormodes}{
\lambda_{1,2} = \beta^m k_m \pm \alpha,\ \lambda_{3,4} = v^mk_m.
}
As in the algebraic gauge, all eigenvalues are purely real. 
The system is strongly hyperbolic, only this time all
modes are non-zero. Thus numerical errors
arising 
from refinement boundary 
interpolations will propagate away.

This is precisely what we observe
in our BHNS simulations and in the following sections we show explicit 
numerical results that demonstrate this. 

\begin{table}
\caption{Summary of comparison runs. 
The third column specifies whether KOD is applied,
and the fourth column indicates whether  
build-up of spurious $B$-field is observed. 
A $\checkmark$ means ``Yes'' and a \ding{55} means ``No''.
  }
\begin{tabular}{cccc}
  \hline  \hline
  Case Name & Gauge & KOD & Spurious $B^i$  \\
  \hline 
  Algebraic  & Algebraic & \ding{55} & $\checkmark$ \\
  Algebraic-KO & Algebraic & $\checkmark$ & $\checkmark$ \\
  Lorenz  & Lorenz & \ding{55} & \ding{55}  \\
\hline\hline
\end{tabular}
\label{table:id}
\end{table}

\section{Numerical Methods}
\label{sec:numerical}

\subsection{Initial data}

BHNS quasiequilibrium initial data are constructed as
described in~\cite{TBFS07b}. The particular BHNS binary chosen in our
comparison of different EM gauges corresponds to Case A of
\cite{elsb09}.  The system consists of an initially nonspinning BH
orbiting about an irrotational NS, with a 3:1 BH:NS mass ratio.  The
NS has a compaction of ${\cal C}=0.145$, and is
constructed using an $n=1$ ($\Gamma=2$) polytropic EOS.

Given that interior NS magnetic field strengths and configurations are unknown,
we choose an initial seed poloidal magnetic field via a vector potential
of the form  
\beqn
  A_i &=& \left( -\frac{y-y_c}{\varpi_c^2}\delta^x{}_i 
+ \frac{x-x_c}{\varpi^2_c}\delta^y{}_i\right) A_\varphi \label{ini:Ai} \\ 
  A_\varphi &=& A_b \varpi^2_c \max (P- P_{\rm cut},0)^{n_b} 
\label{ini:Aphi}
\eeqn
where $(x_c,y_c,0)$ is coordinate location of the center of mass of the NS, 
$\varpi^2_c=(x-x_c)^2+(y-y_c)^2$, and $A_b$, $n_p$ and $P_{\rm cut}$ are free 
parameters. The cutoff pressure parameter $P_{\rm cut}$ confines the B-field
within the neutron star to $P>P_{\rm cut}$. The parameter $n_b$ 
determines the degree of central condensation of the magnetic field. 
Similar profiles of initial magnetic fields have been used in numerical 
simulations of magnetized accretion disks (see, e.g.~\cite{dhk03,mg04}) and
magnetized compact binaries (see, e.g.~\cite{ahllmnpt08,grb11,cabllmn10}). 
 We set $P_{\rm cut}$
to be 4\% of the maximum pressure and $n_b$=2. The parameter $A_b$ controls 
the strength of the initial magnetic field and can be characterized by 
the maximum magnetic field inside the NS, which we set here
to $B=1.3\times 10^{16} G$ (assuming the NS mass is $1.4M_{\odot}$).
Such seed magnetic fields are too weak to significantly perturb the
initial quasiequilibrium NS, and thereby do not contribute to any 
significant initial gravitational field constraint violations.

\subsection{Evolution algorithms}
\label{sec:num_metric_hydro}

Our methods for evolving the metric and the MHD equations
are described in \cite{elsb09} and for this reason are not presented
here. This section focuses solely on our treatment of the
evolution of the vector potential.

\begin{table*}
\caption{Grid configuration used in our simulations. $N_{\rm
    AH}$ denotes the number of grid points covering the diameter of
    the apparent horizon initially, and $N_{\rm NS}$
    denotes the number of grid points covering the smallest diameter
    of the NS initially.}
\begin{tabular}{cccc}
  \hline  \hline
  Grid Hierarchy (in units of $M$)$^{(a)}$
  & Max.~resolution & $N_{\rm AH}$ & $N_{\rm NS}$ \\
  \hline    
(203.7, 101.8, 25.46, 12.73, 6.365, 3.183, 1.591) & $M/31.4$ & 40 & 82 \\
\hline\hline
\end{tabular}
\begin{flushleft}
$^{(a)}$ There are two sets of nested refinement boxes: one centered
  on the NS and one on the BH.  This column specifies the half side length
  of the refinement boxes centered either on the BH or the NS. 
\end{flushleft}
\label{table:GridStructure}
\end{table*}

The magnetic induction equation is evolved via the 4-vector 
potential using Eqs.~(\ref{eq:Aevol}) and (\ref{eq:Phievol}). 
$A_i$ and $B^i$ are staggered as in~\cite{ELS2010}, and
$\Phi$ exists on the $(i^+,j^+,k^+)$ staggered grid [all other 
hydrodynamic, BSSN, lapse and shift variables are stored at $(i,j,k)$], 
where $(i^+,j^+,k^+) = (i+1/2,j+1/2,k+1/2)$.
The term $-\partial_j (\beta^j \sqrt{\gamma}\, \Phi)$ in 
Eq.~(\ref{eq:Phievol}) is treated using a second-order upwind
scheme, and Eq.~(\ref{eq:Aevol}) is handled using the finite-volume
equations similar to Eqs.~(63)--(65) in~\cite{ELS2010}. For example,
we evolve $\hat{A}_x$, which is staggered at $(i,j^+,k^+)$, by 
\beqn
  \partial_t (\hat{A}_x)_{i,j^+,k^+} 
&=& -\hat{\cal{E}}^x_{i,j^+,k^+} -\frac{1}{\Delta x} 
[ (\alpha \Phi - \beta^j A_j)_{i^+,j^+,k^+} \cr
&& -(\alpha \Phi - \beta^j A_j)_{i^-,j^+,k^+}] ,
\label{eq:dtAx}
\eeqn
where $i^-=i-1/2$, $\hat{A}_x$ is the line-averaged $A_x$ defined in 
Eq.~(57) of~\cite{ELS2010}, ${\cal E}^i = \epsilon^i{}_{jk} B^j v^k$ 
and $\hat{\cal{E}}^x$ is the line-averaged ${\cal E}^x$ defined 
in Eq.~(49) of~\cite{ELS2010}. In~\cite{ELS2010} and~\cite{zbl03}, 
only the first term on the right-hand side of Eq.~(\ref{eq:dtAx}) 
is present because the algebraic gauge Eq.~\eqref{algebraic}
is imposed. The term ${\cal E}^x_{i,j^+,k^+}$ is 
computed using the HRSC scheme described in~\cite{ELS2010}
and~\cite{zbl03}. Values of $\alpha$, $\beta^j$ and $A_i$ at the staggered points 
$(i^\pm,j^+,k^+)$ on the right-hand side of Eq.~(\ref{eq:dtAx}) 
are computed by simple averaging. Similar equations can be derived 
for $\hat{A}_y$ and $\hat{A}_z$.

$A_i$ is staggered to ensure that 
$B$ fields obtained by taking the curl operator on $A_i$ [Eqs.~(60)--(62)
in~\cite{ELS2010}] are numerically equivalent to those obtained from the
standard, staggered-$B$ constrained transport scheme~\cite{eh88}.
It is easy to prove that the 
additional terms in Eq.~(\ref{eq:dtAx}) and the corresponding terms in 
$A_y$ and $A_z$ equations cancel exactly after applying the 
curl operator. The resulting numerical values of $B^i$ are thus 
gauge-invariant in unigrid simulations. We have confirmed numerically 
that this is indeed the case. However, in simulations with an AMR grid, 
since interpolations on $A_i$ are performed between refinement levels, 
values of $A_i$ are not the same in different EM gauges at the refinement 
boundaries. The resulting $B$ field at the refinement boundaries are 
also different in general but should converge to a unique, true solution 
with increasing resolution in any gauge.

\subsection{Recovery of primitive variables}
\label{sec:inversion}

At each timestep, 
the ``primitive variables'' 
$\rho_0$, $P$, and $v^i$ are recovered from the ``conservative''
variables  $\rho_*$, $\tilde{\tau}$, and $\tilde{S}_i$. 
We perform the inversion by numerical solving two nonlinear
equations via the Newton-Raphson method as described in~\cite{ngmz06},
using the public code developed by Noble et al.~\cite{harmsolver}.

Sometimes numerical errors may drive ``conservative'' variables to
unphysical values, resulting in unphysical
primitive variables after inversion (e.g.\ negative pressure or even
complex solutions). This usually happens in the low-density 
``atmosphere'' or deep inside the BH interior where high-accuracy 
evolutions are difficult to maintain. Various techniques have been
proposed to handle inversion failures (see, e.g.~\cite{bs2011}). 
Our approach imposes constraints on the conservative 
variables to reduce inversion failures (for details see
\cite{elpb11}).

\begin{figure*}[h]
\centering
\subfigure{\includegraphics[width=0.3\textwidth]
{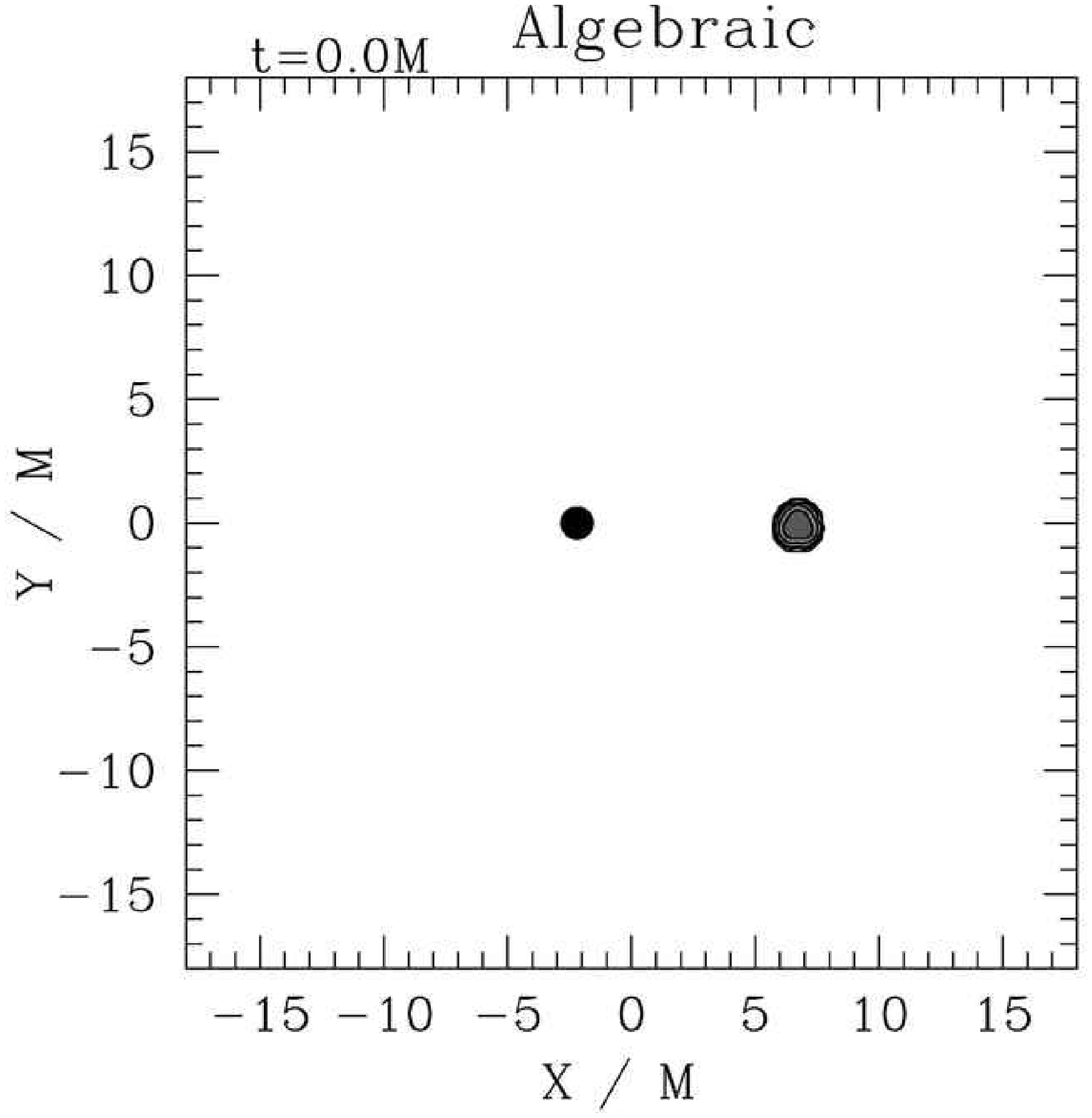}}
\subfigure{\includegraphics[width=0.3\textwidth]
{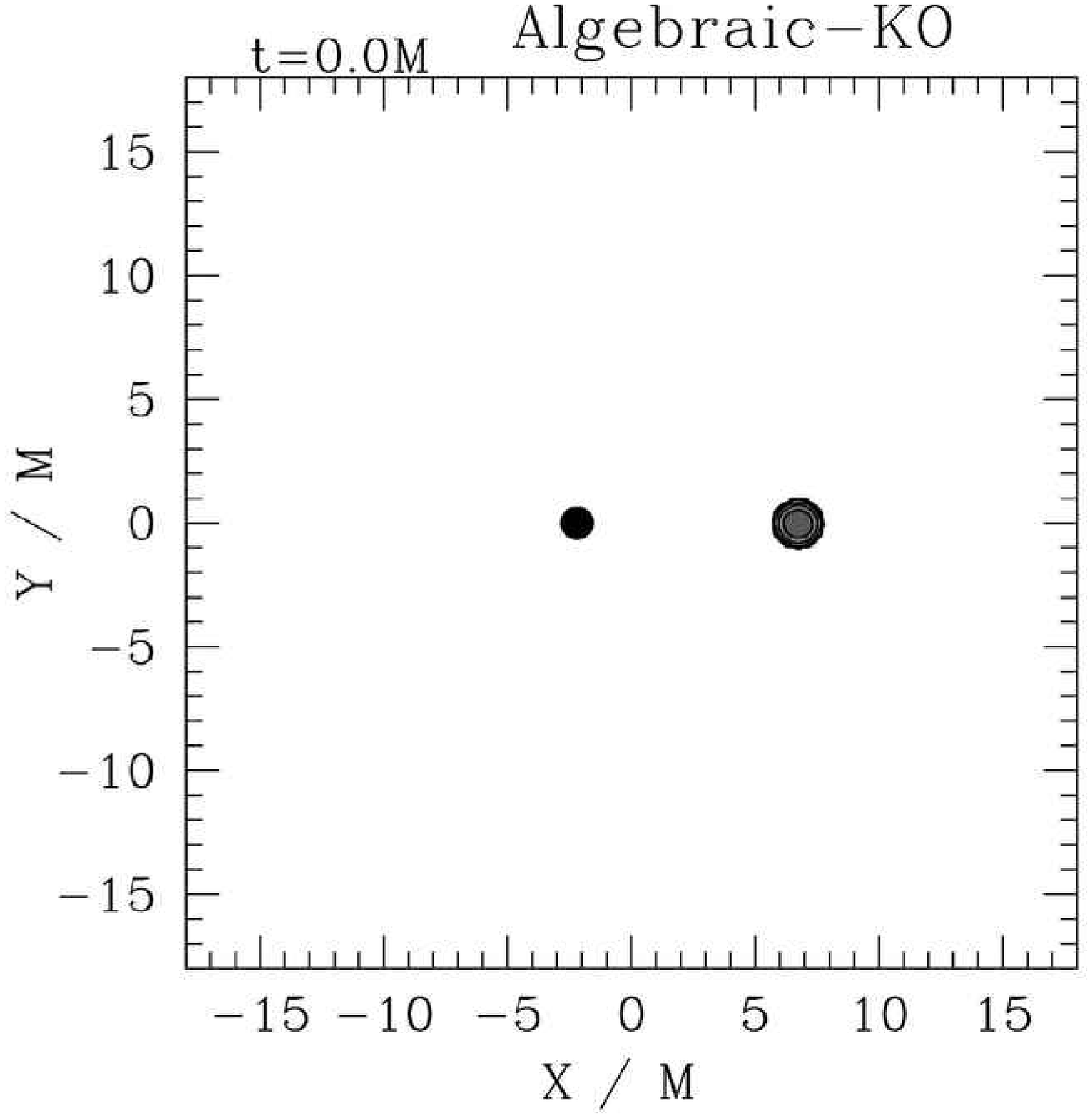}}
\subfigure{\includegraphics[width=0.3\textwidth]
{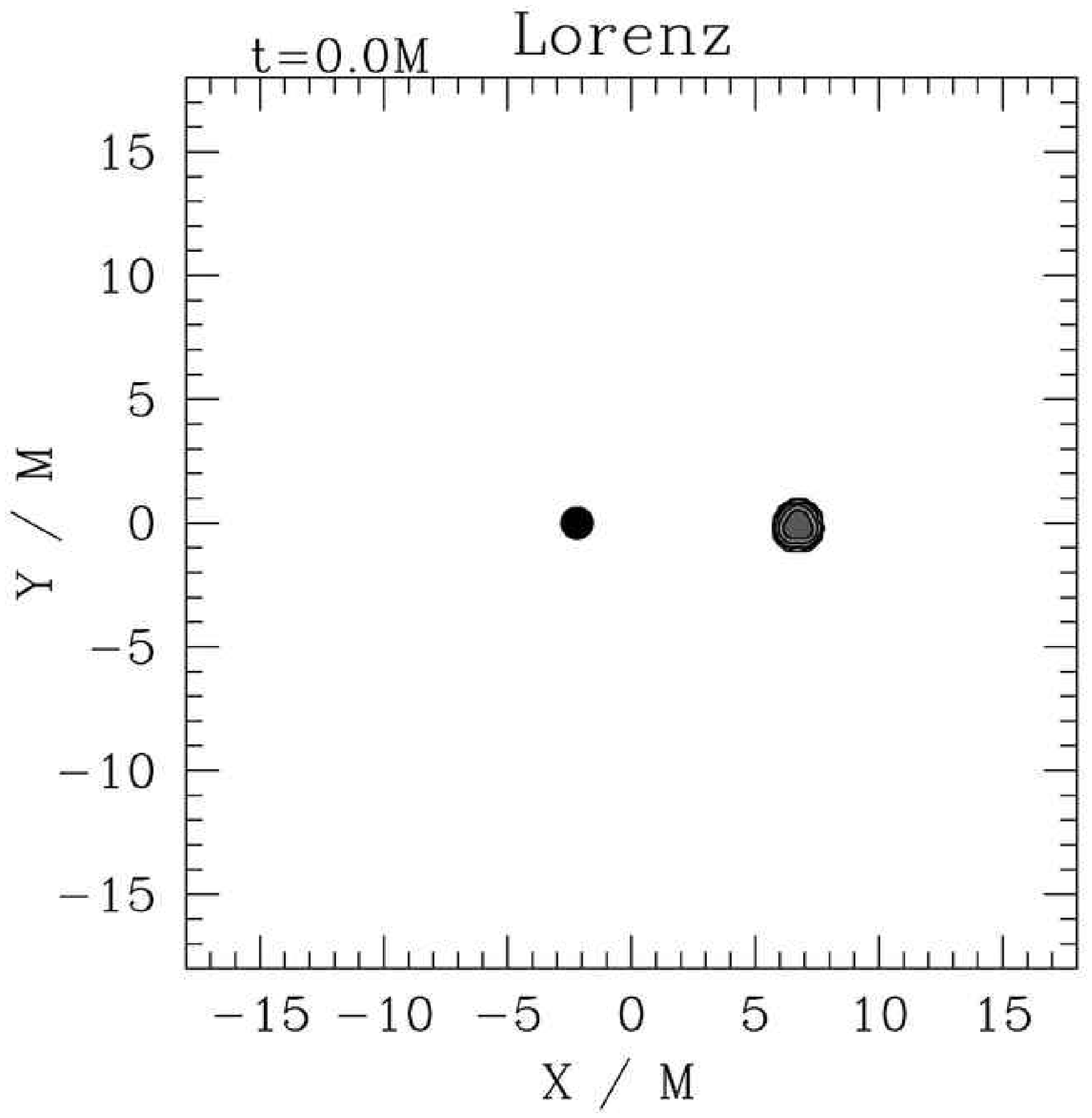}}
\subfigure{\includegraphics[width=0.3\textwidth]
{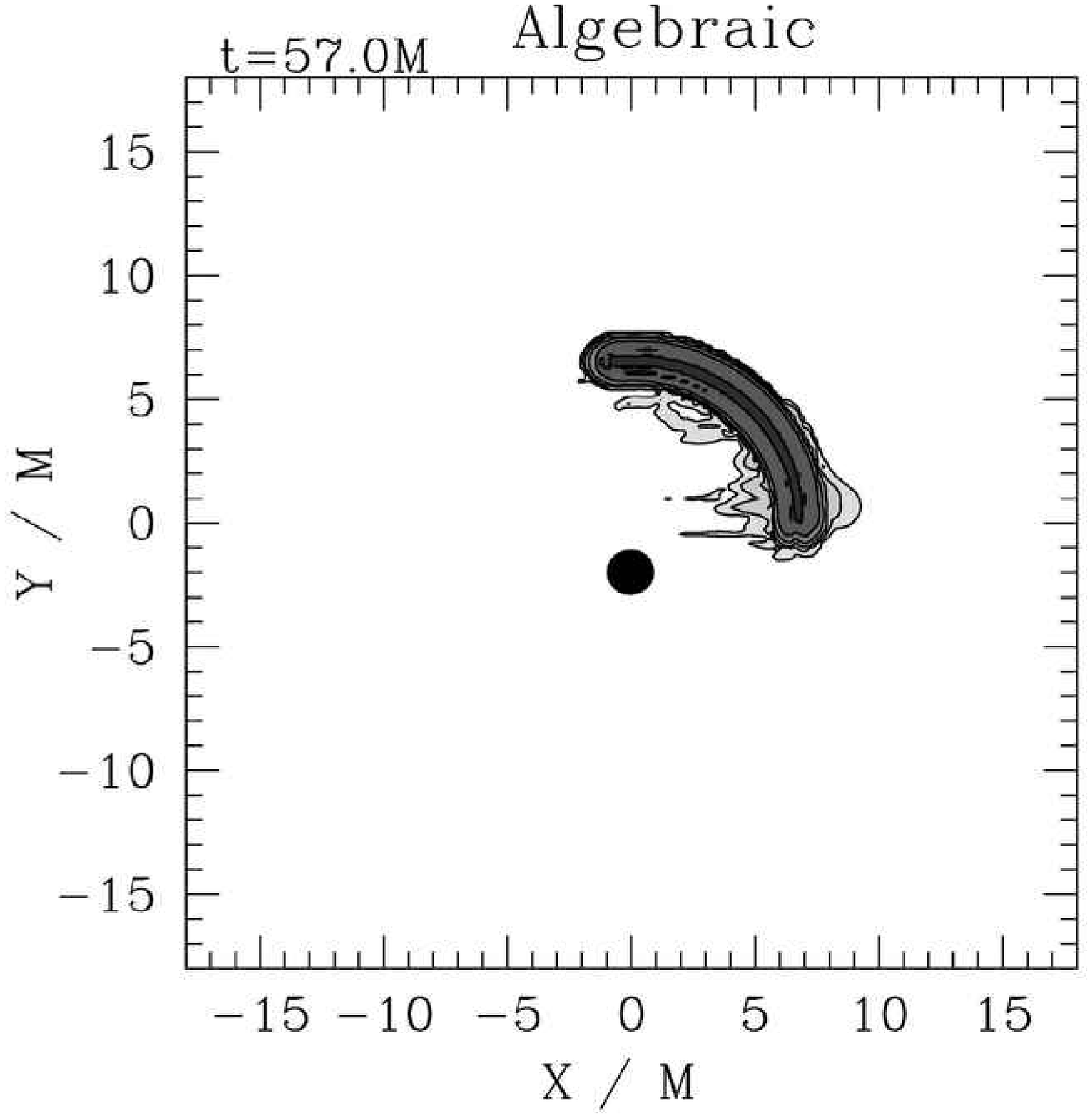}}
\subfigure{\includegraphics[width=0.3\textwidth]
{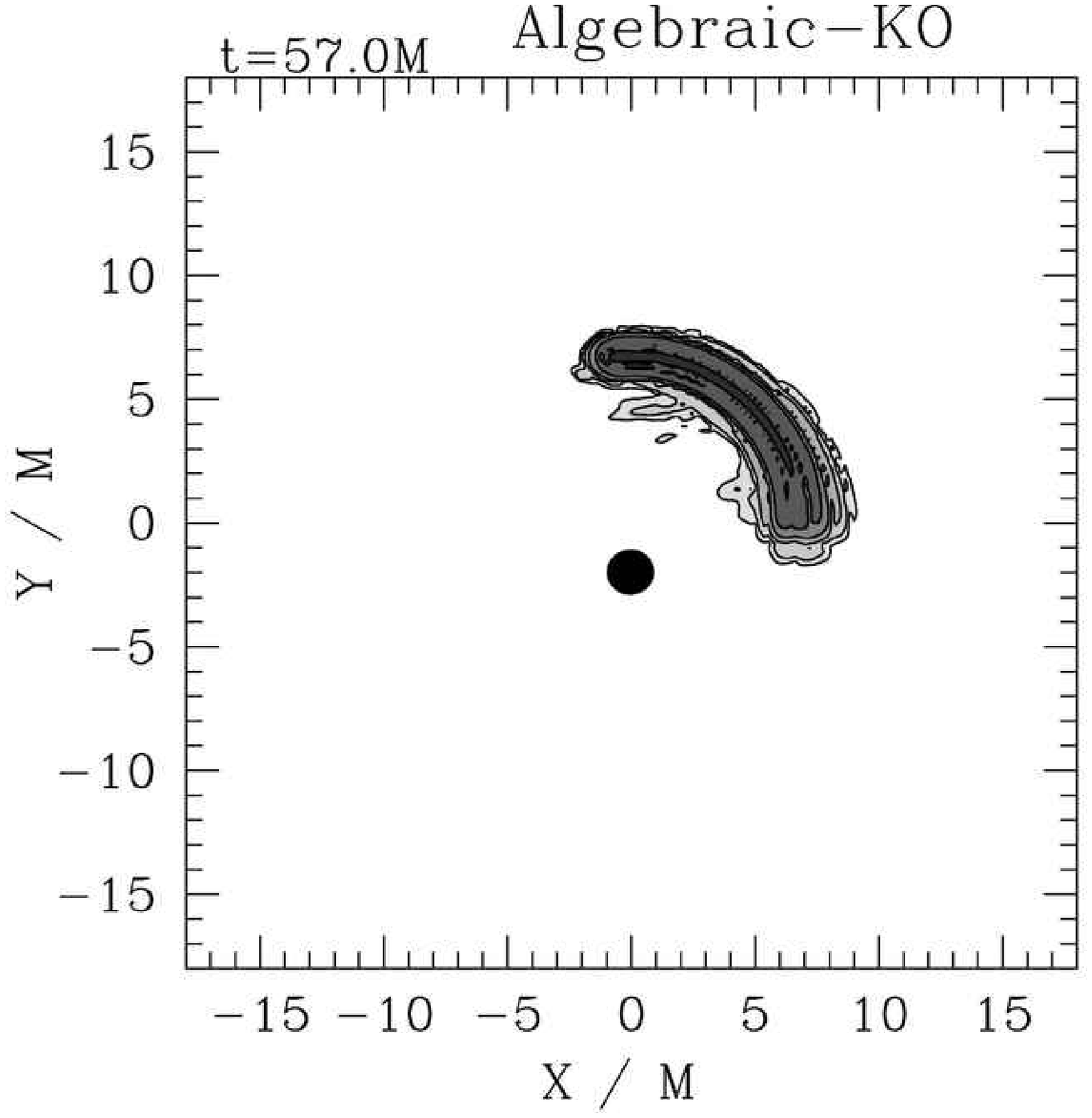}}
\subfigure{\includegraphics[width=0.3\textwidth]
{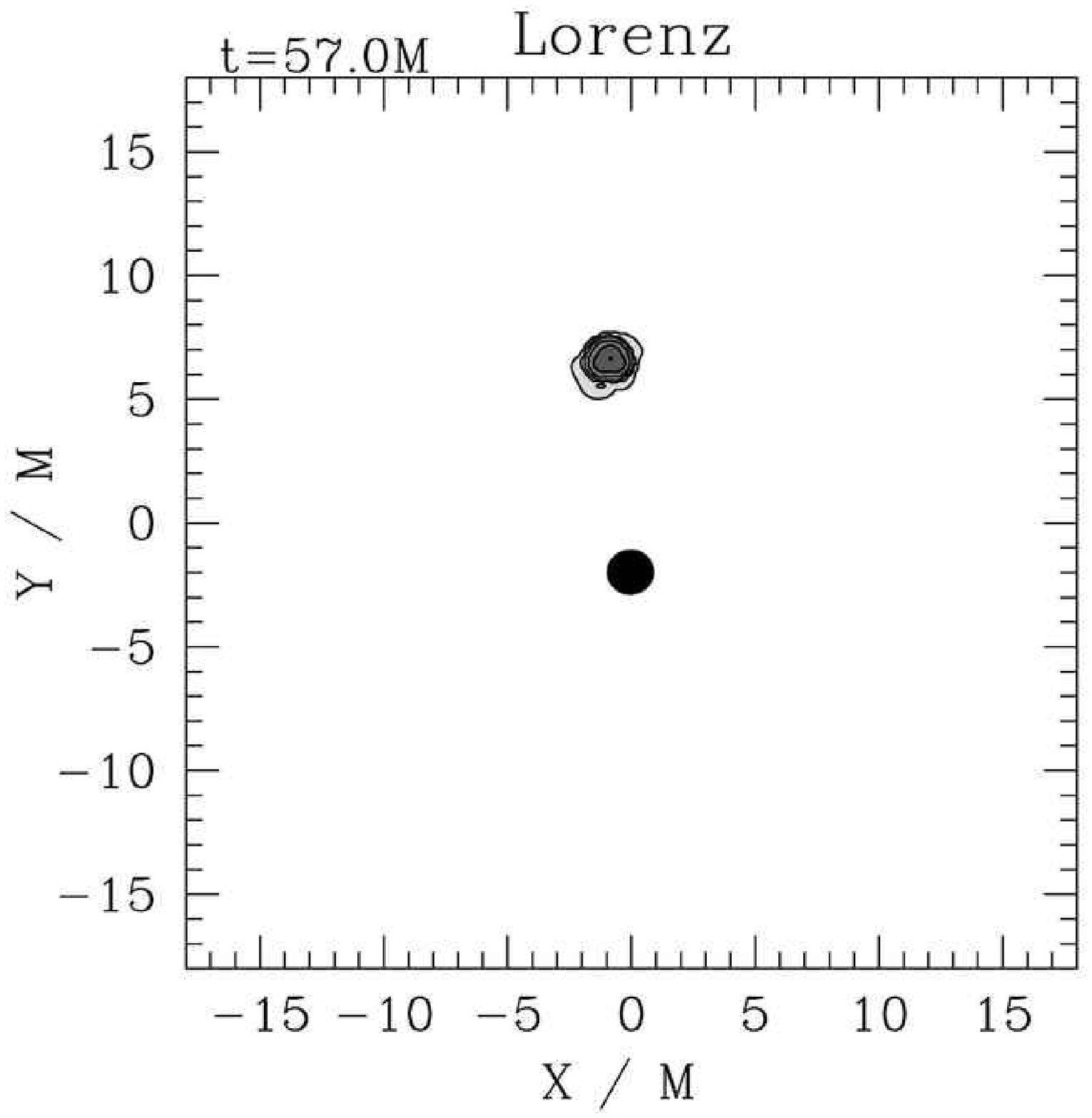}}
\subfigure{\includegraphics[width=0.3\textwidth]
{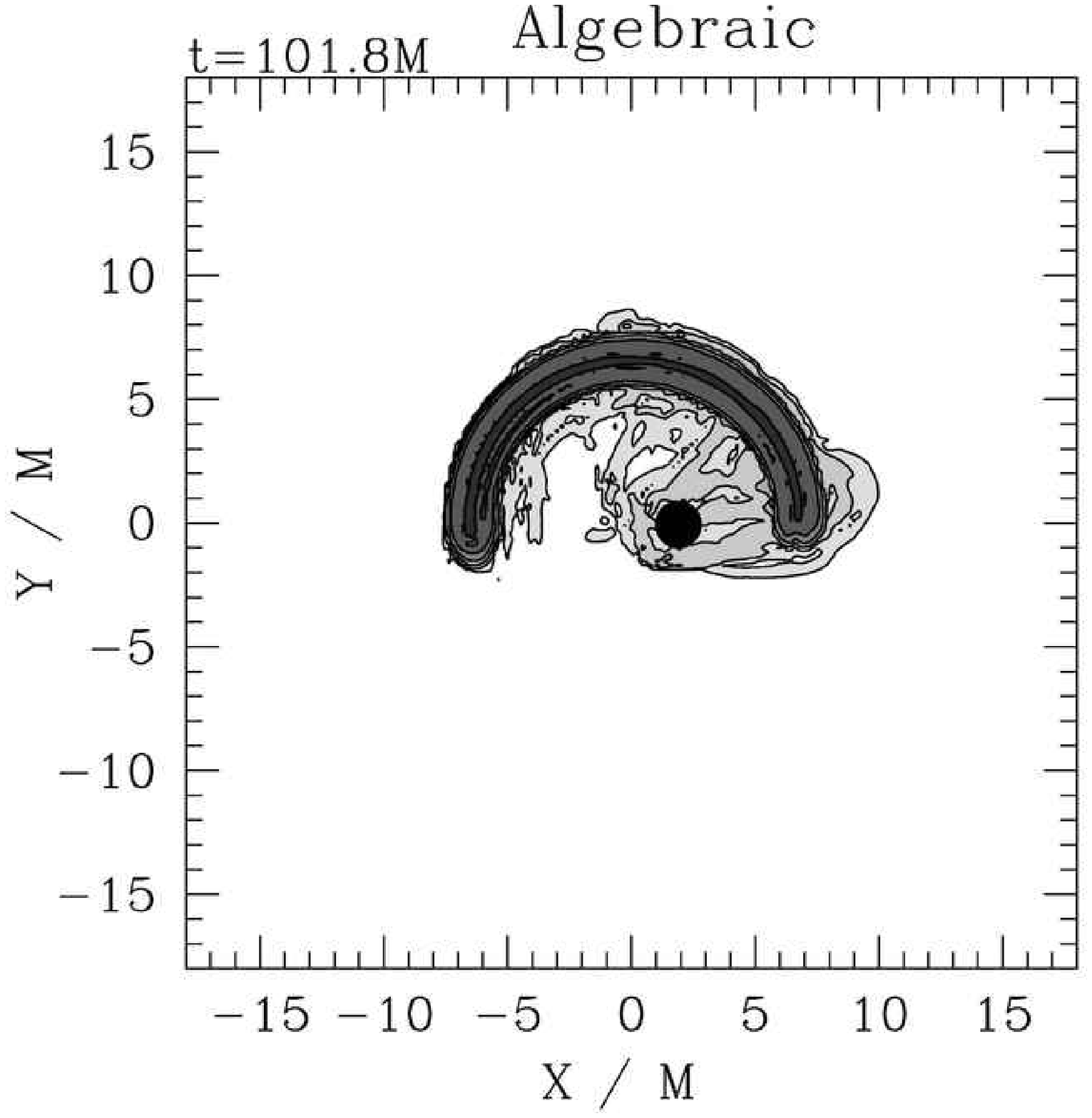}}
\subfigure{\includegraphics[width=0.3\textwidth]
{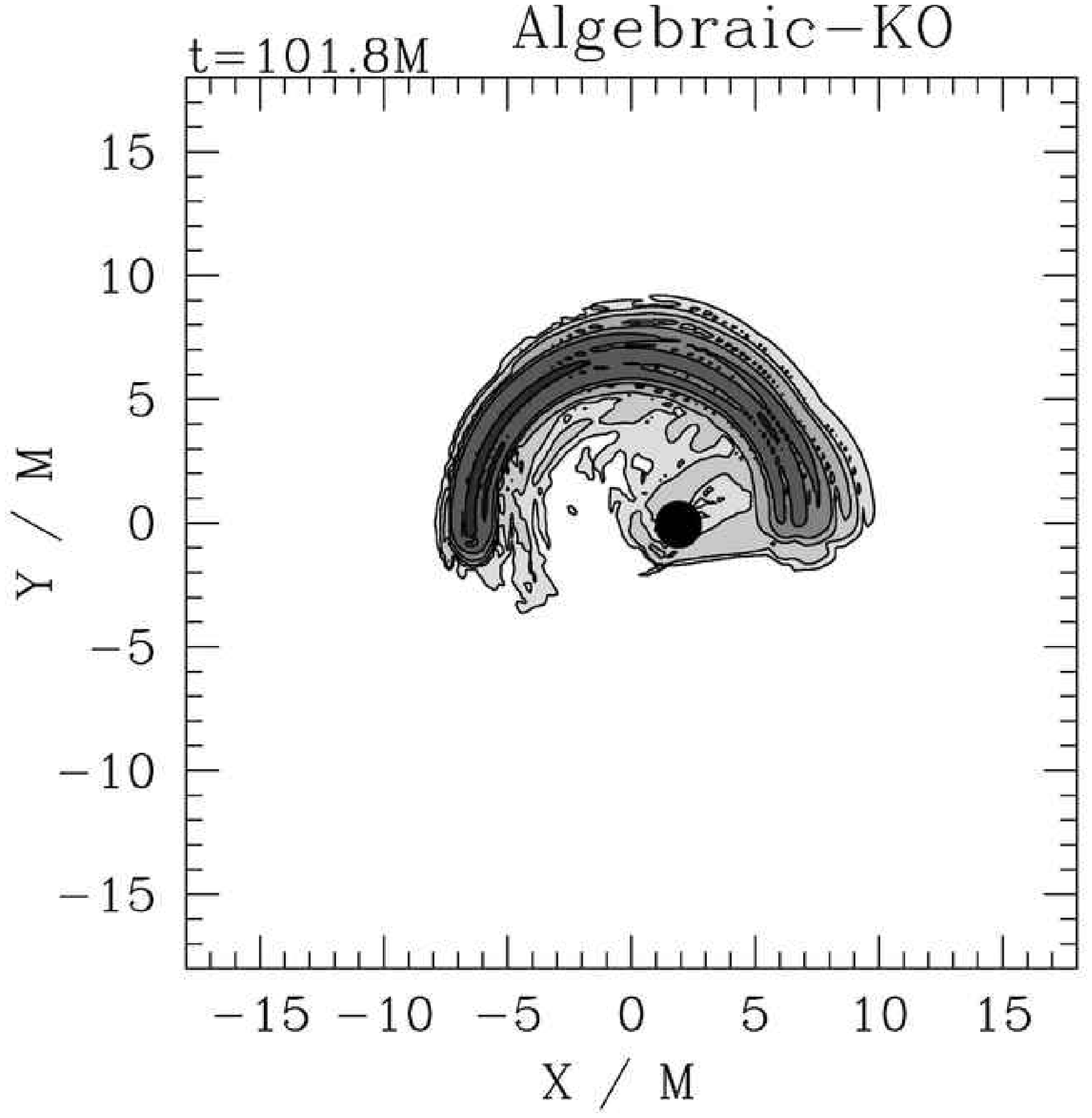}}
\subfigure{\includegraphics[width=0.3\textwidth]
{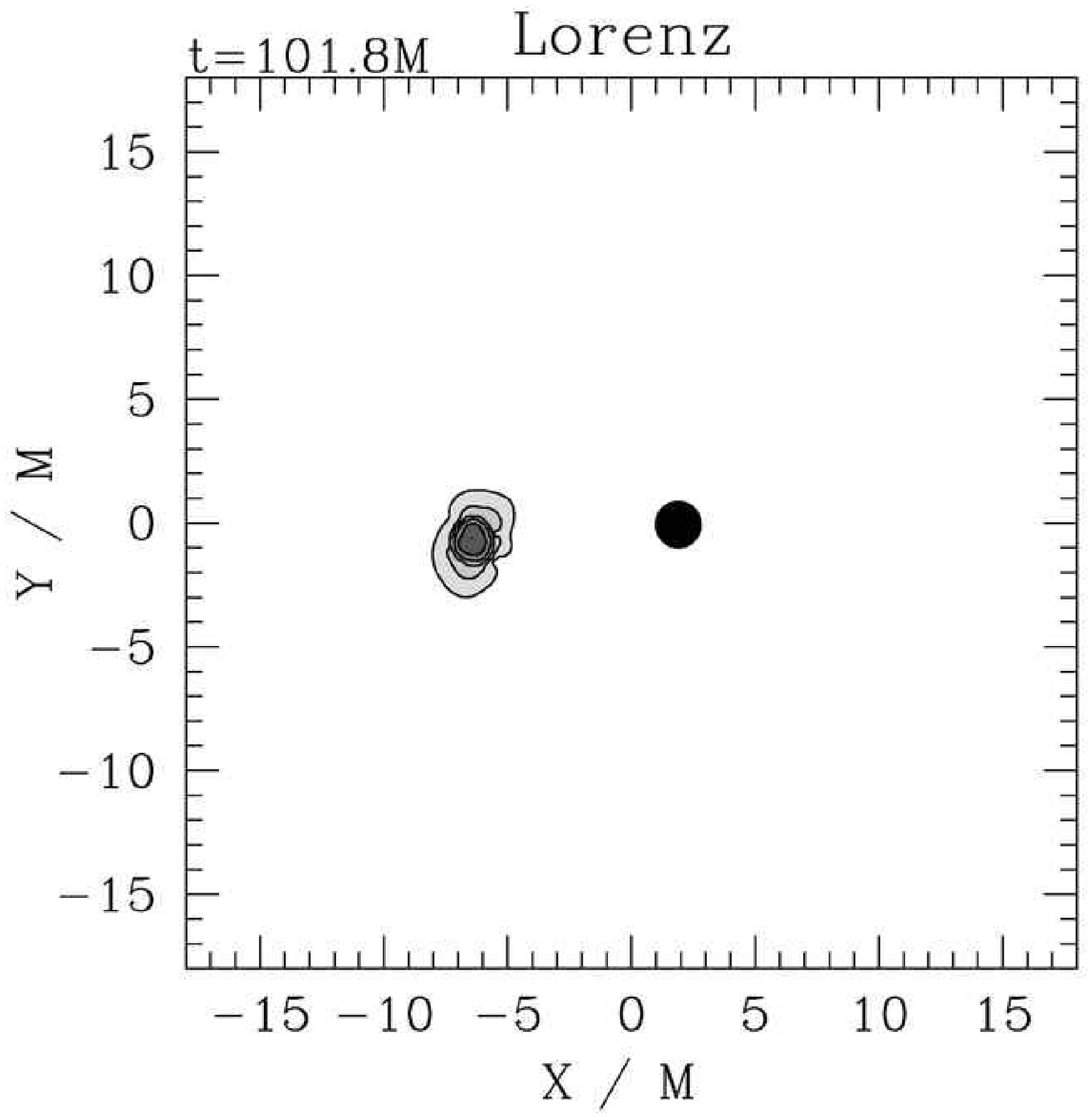}}
\subfigure{\includegraphics[width=0.3\textwidth]
{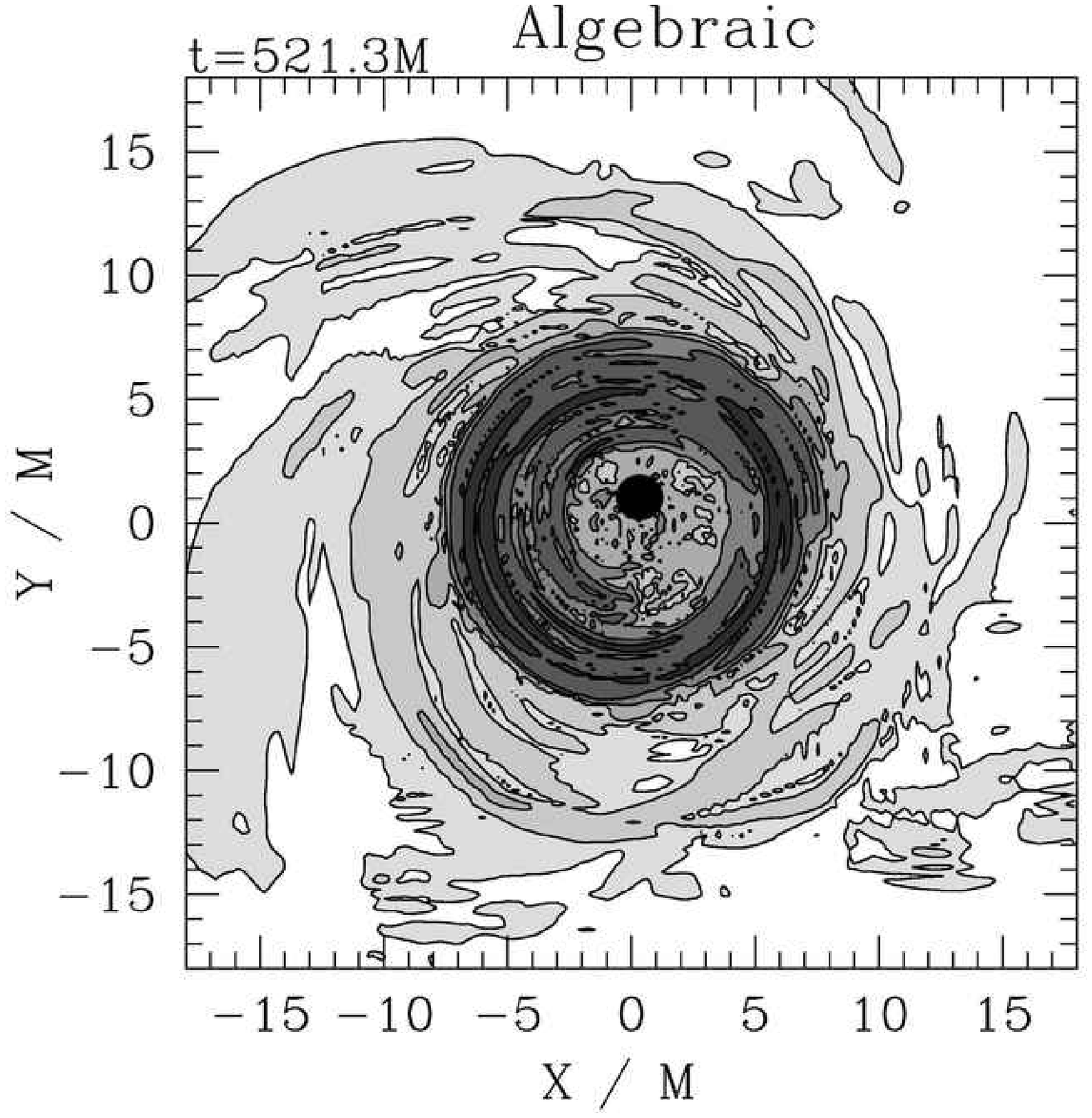}}
\subfigure{\includegraphics[width=0.3\textwidth]
{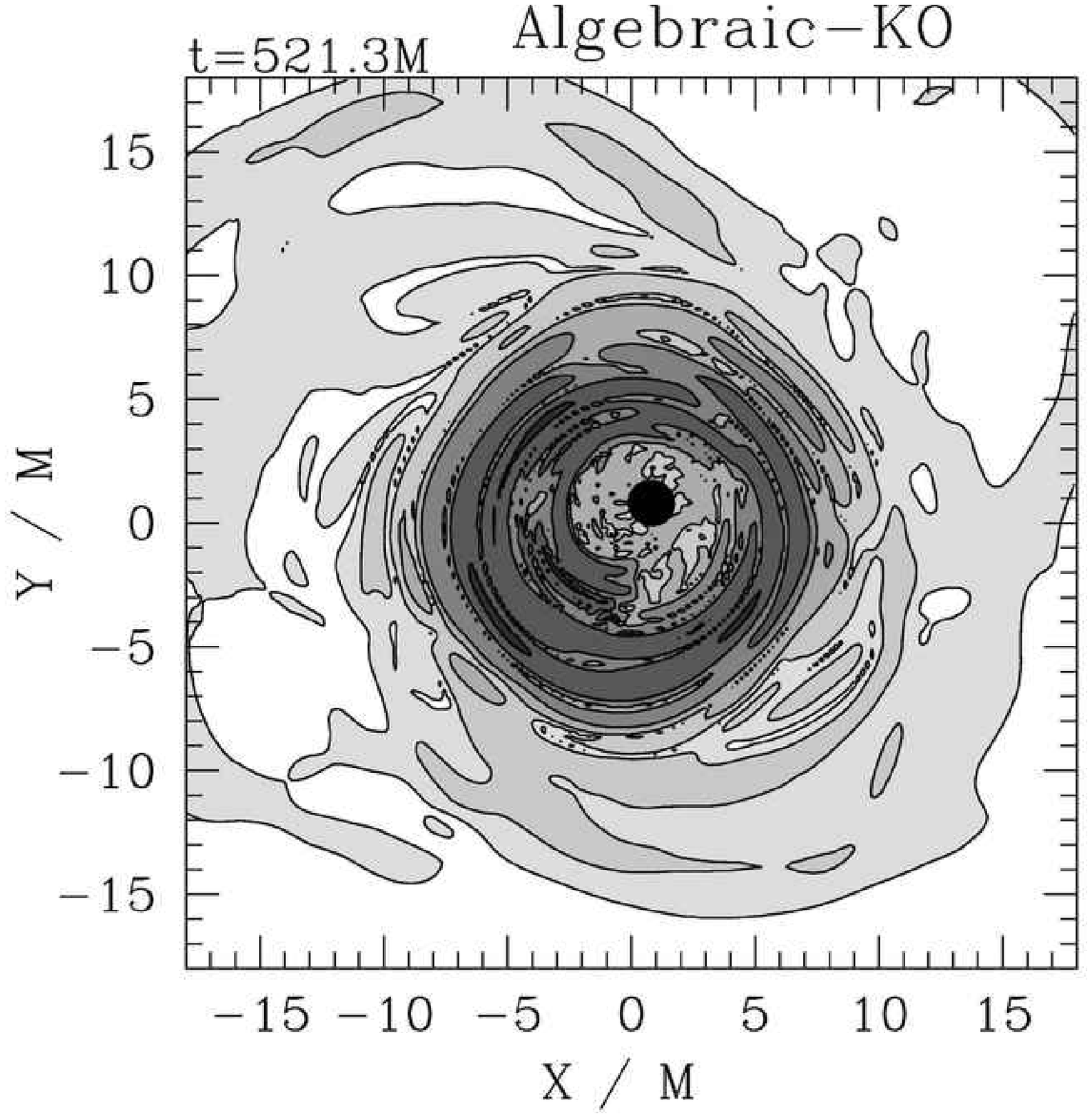}}
\subfigure{\includegraphics[width=0.3\textwidth]
{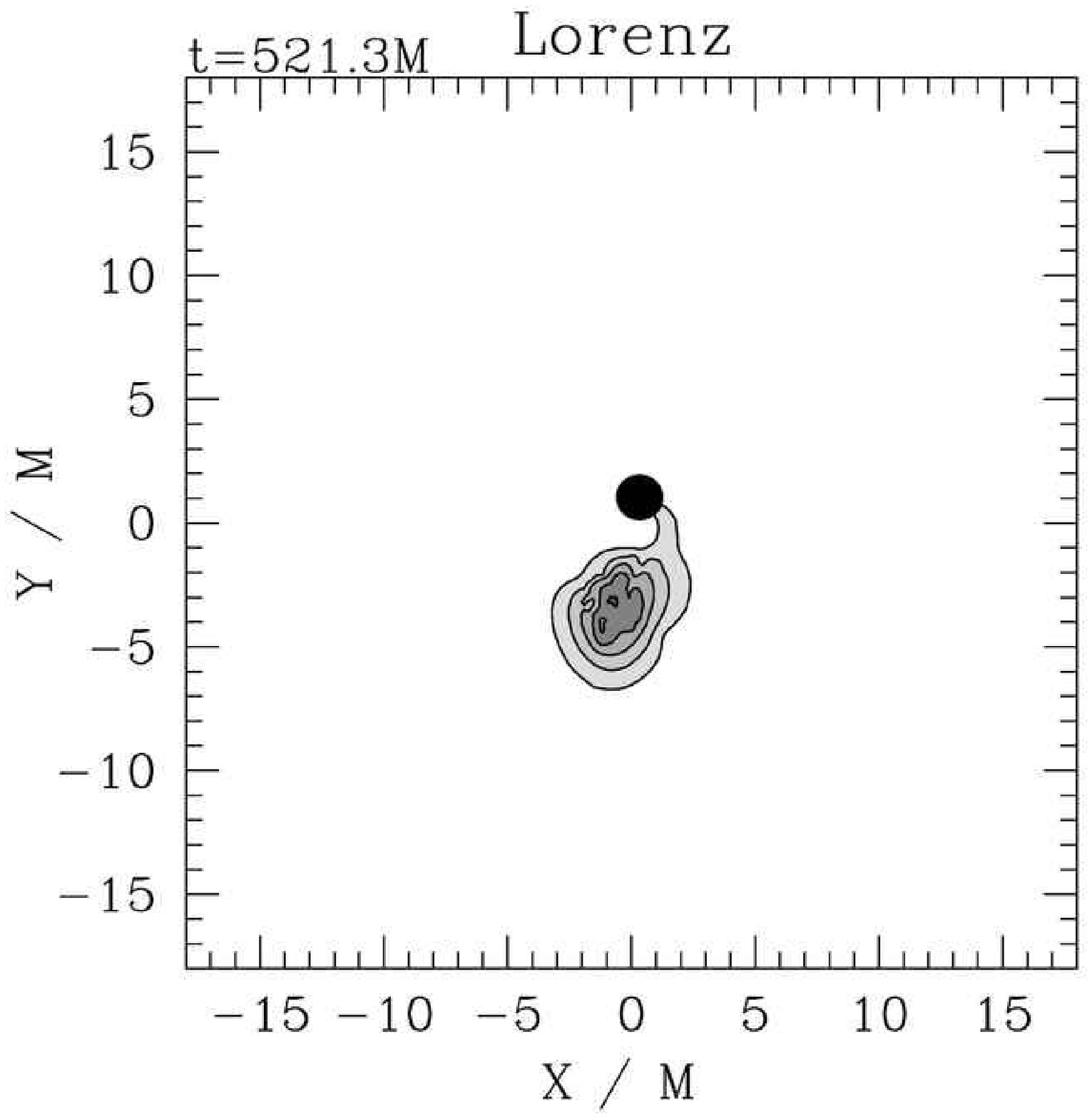}}
\caption{
Snapshots of $|\mathbf{A}|$ (the magnitude of $A_i$) profiles
in the orbital plane of a BHNS binary plotted at selected times 
according to $|\mathbf{A}|=10^{-3-0.6i},\ i=0,\ldots, 5$,
with darker greyscaling for stronger $|\mathbf{A}|$. Notice the
$A_i$ trail left behind the orbiting NS in the algebraic 
gauge with or without KOD, and the absence of such trail 
in the Lorenz gauge. This trail is an immediate 
consequence of the existence of a static mode in the
algebraic gauge. When the refinement boxes
intersect the $A_i$ trail, spurious magnetic fields are generated. 
The filled black circled near the center denotes the BH's horizon.
\label{fig:LorenzvsnonLorenzA2}
}
\centering
\end{figure*}


\begin{figure*}
\centering
\subfigure{\includegraphics[width=0.3\textwidth]
{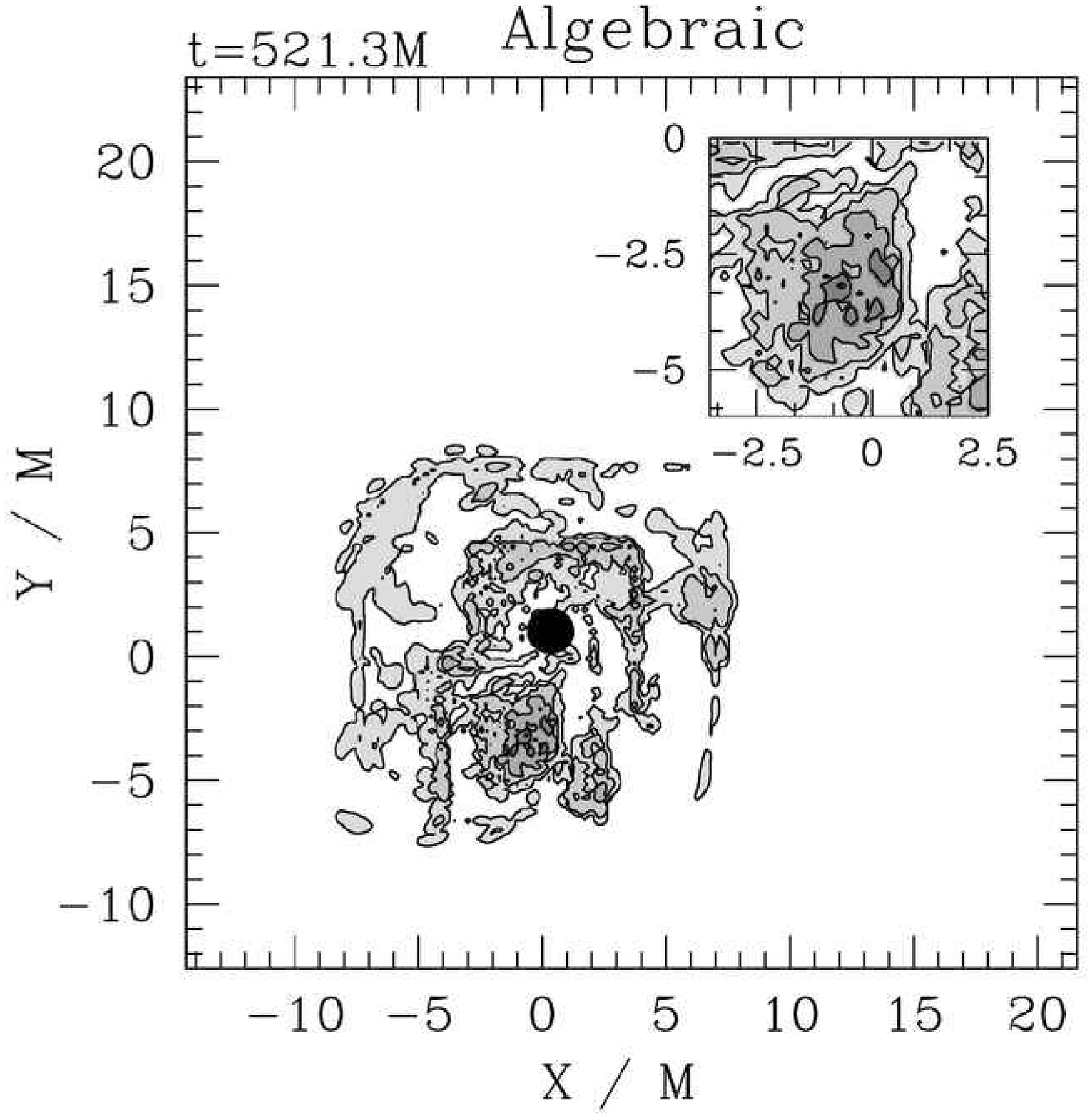}}
\subfigure{\includegraphics[width=0.3\textwidth]
{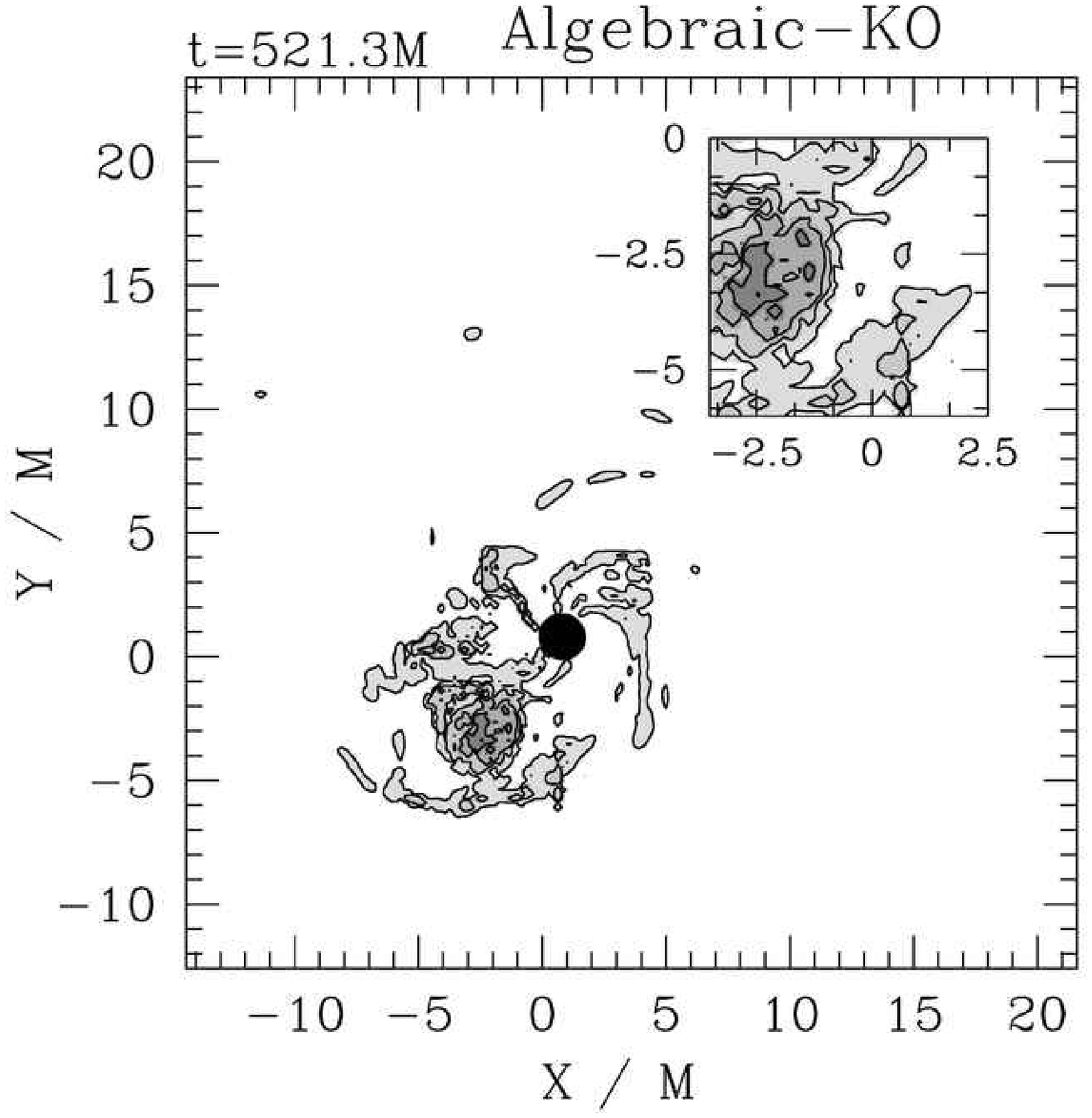}}
\subfigure{\includegraphics[width=0.3\textwidth]
{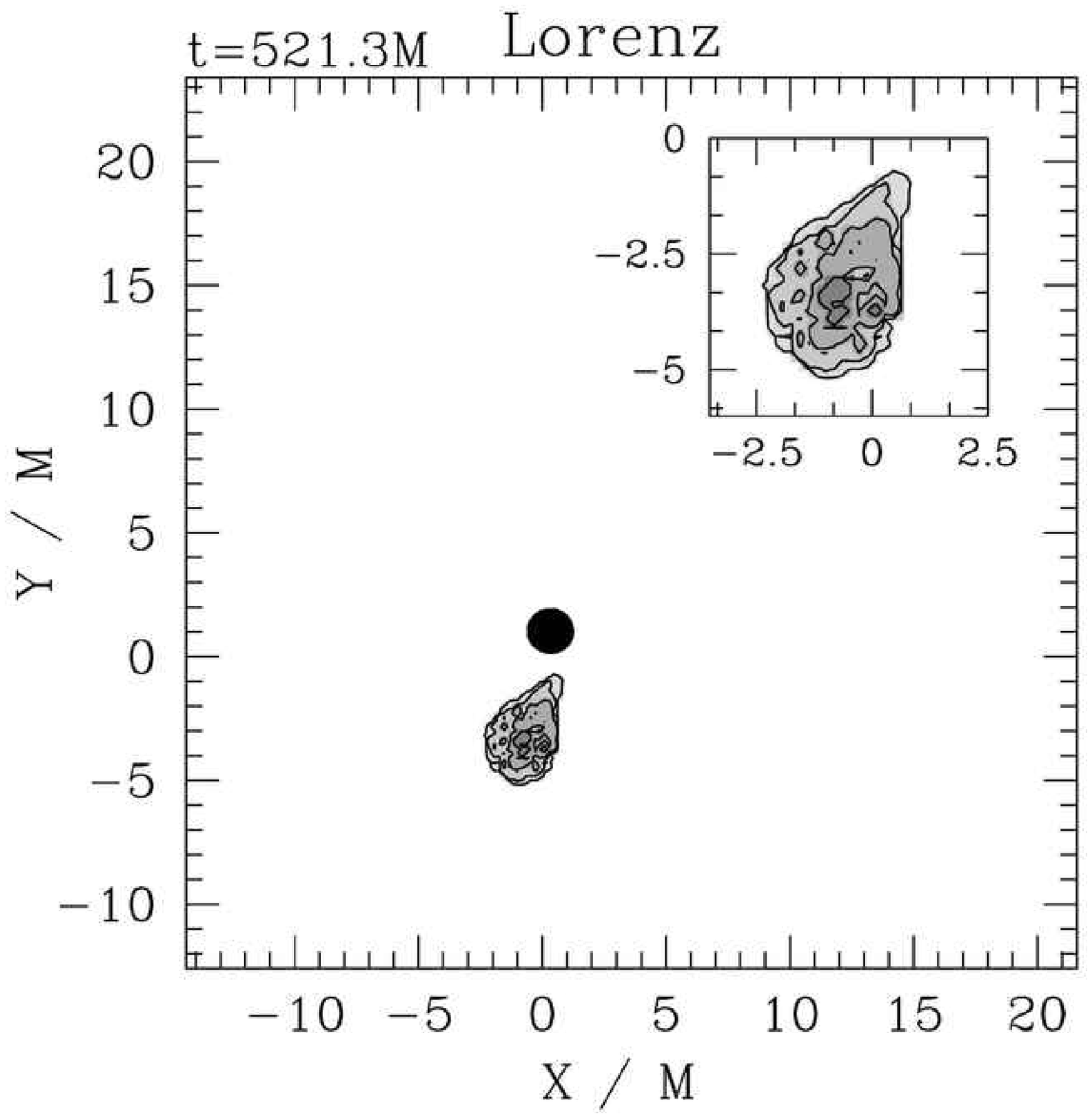}}
\caption{
Orbital-plane snapshots of the comoving magnetic energy density
($b^2$) just before NS tidal disruption. $b^2$ 
profiles are plotted according to
$b^2=10^{-4.49-1.4i}/M^2,\ i=0,\ldots, 5$, with 
darker greyscaling for larger values. 
The insets zoom in on the NS and demonstrate that the 
B-field structure is maintained in the outer layers of the NS when using
Lorenz gauge, whereas it is blown away at this time when the algebraic
gauge is used with or without KOD. Note also the spurious B-fields
that have appeared at refinement level boundaries using the algebraic gauge. 
These snapshots clearly demonstrate that the algebraic gauge with or
without KOD generates spurious magnetic fields and destroys 
the B-field structure of the NS.
\label{fig:LorenzvsnonLorenzb2}
}
\centering
\end{figure*}

\section{Results}
\label{sec:results}

\subsection{Cases and grid structure}

Table~\ref{table:id} summarizes all simulations presented here.
The only parameters varied in the BHNS simulations shown
here are the EM gauge condition and the application of
fourth-order KOD to the $A_i$ evolution of the 
form $\epsilon/16 (\Delta x^3 \partial_x^{(4)}+
\Delta y^3 \partial_y^{(4)}+\Delta z^3 \partial_z^{(4)})f$
(with strength parameter set to 0.1). Here $f$ denotes the vector potential. 
Note that this KOD term converges to zero at third-order in the grid 
spacing, so that it does not alter the convergence properties 
of our EM evolution scheme.
One simulation adopts the Lorenz gauge without KOD
(named ``Lorenz''), and two simulations adopt the 
algebraic gauge -- one with and one without KOD, called
``Algebraic-KO'' and ``Algebraic'', respectively.
The grid structure used in our simulations is presented in
Table~\ref{table:GridStructure} and is the same for all cases. 
Our main findings are summarized in Figs.~\ref{fig:LorenzvsnonLorenzA2} and 
\ref{fig:LorenzvsnonLorenzb2}.

\subsection{Algebraic gauge}

Figure~\ref{fig:LorenzvsnonLorenzA2} demonstrates that in the
algebraic gauge (left two columns), there is an eigenmode of the
$A_i$-field which does not propagate (as derived in Eq.~\eqref{algmodes})
leading to a path of $A_i$ fields that trails
the orbiting NS. Two nested sets of adaptive-mesh
refinement boxes surround and comove with the NS and BH, with lower
resolution at larger distances from these objects (see
Table~\ref{table:GridStructure}).  When refinement levels cross
the $A_i$ trail, the $A_i$ fields are filtered to lower or
higher frequencies.  This filtering leads to the appearance of spurious,
strong magnetic fields on refinement box boundaries (see
Fig.~\ref{fig:LorenzvsnonLorenzb2}). When the spurious, strong
magnetic fields appear, the magnetic energy outside  
the horizon jumps by about 5\% after about half an orbit -- a strong effect.  
Kreiss-Oliger dissipation helps to damp the spuriously generated magnetic 
fields, resulting in an initial net {\it drop} of magnetic energy outside the
horizon by 5\% after about half an orbit.   Corresponding 
to this drop in magnetic energy, KOD gradually smooths the initially steep
$A_i$-field gradient in the NS, ultimately destroying the NS's magnetic
field structure (inset of middle frame in Fig.~\ref{fig:LorenzvsnonLorenzb2}). 
Just before tidal disruption
the magnetic energy outside the horizon in the 
Algebraic and the AlgebraicKO cases is about 17\% and 77\% higher 
than the magnetic energy in the Lorenz case, respectively.
Moreover, when KOD is turned on, the $b^2$ contours lag behind 
the corresponding $b^2$ contours when employing either the algebraic 
gauge without KOD or the Lorenz gauge (compare the insets 
in Fig.~\ref{fig:LorenzvsnonLorenzb2}).

\subsection{Lorenz gauge}

In contrast to the algebraic gauge, there are no zero-speed modes 
in the Lorenz gauge (see Eq.~\eqref{lormodes}), and no
$A_i$ field trail appears behind the NS (third-column
plots in Fig.~\ref{fig:LorenzvsnonLorenzA2}).  
Immediately after the simulation begins, an initial burst of
an $A_i$ gauge wave
(with a negligibly small $B$ field associated with it)
emanates from the NS, propagates outwards and is partially
absorbed by the BH. This 
wave travels at the speed of light, as expected for the gauge modes
 $\lambda_{1,2}$ in Eq. \eqref{lormodes}. 
As the NS orbits the BH, the $A_i$ field comoves with the plasma.
Therefore, strong, spurious magnetic fields 
at refinement boundaries do not appear (right frame in
Fig.~\ref{fig:LorenzvsnonLorenzb2}). Also, since KOD is not used, the
original magnetic field  
structure of the NS is better-maintained 
(inset of right panel in Fig.~\ref{fig:LorenzvsnonLorenzb2})
than in the algebraic gauge simulations.

\section{Summary}
\label{sec:discussion}

In developing an GRMHD algorithm that maintains $\mathbf{\nabla\cdot B} = 0$
and is compatible with the moving puncture technique, we have focused on CT
schemes.  Such schemes guarantee that $\mathbf{\nabla\cdot B} = 0$ is
maintained to truncation error.  In \cite{ELS2010}, we developed an
AMR-compatible CT scheme, in which the magnetic induction equation is
recast as an evolution equation for the magnetic vector potential. The
divergence-free magnetic field is then computed via the curl of
the vector potential.  Unlike the
magnetic field, the vector potential is not constrained,
and so any interpolation scheme can be used during prolongation
and restriction, thus enabling its use with any AMR algorithm.
In addition, given that the vector potential does not uniquely
determine the magnetic field, there is an extra gauge degree
of freedom that can be used to one's advantage. 

In \cite{ELS2010} we performed several tests on our new AMR-compatible 
CT scheme. Using an algebraic EM gauge condition
we found that our scheme works well even in black-hole spacetimes. 
Hence our scheme is compatible with the moving puncture technique.
However, when we use this EM gauge condition in the evolution of
magnetized binary BHNSs, stable, long-term simulations are not
possible.

In this paper, we introduced a new EM gauge condition, 
the Lorenz gauge, which allows for stable, 
long-term GRMHD evolutions of binary BHNSs. 
Performing an eigenvalue analysis we found
that static EM modes are present when adopting the algebraic gauge, 
whereas all modes are propagating when the Lorenz gauge is adopted. 

Using magnetized BHNS simulations, we confirmed the expected behavior
implied by the eigenmode analysis.
As the magnetized NS orbits the BH, static modes in the algebraic
gauge lead to a trail of nonzero $A_i$ left behind the NS.
When such static modes from the algebraic gauge are interpolated at
refinement boundaries, spurious B-fields are generated which remain on
the grid.  This spurious effect contaminates the solution and is
amplified in time, forcing the simulations to terminate shortly after the
NS tidal disruption.  In contrast to the algebraic gauge, when the
Lorenz gauge is adopted, such spurious effects are quickly propagated
away to the boundary and leave the computational domain.
Overall we find that the Lorenz gauge provides a major improvement  
over the algebraic gauge for magnetized BHNS simulations. 
In a forthcoming paper~\cite{elpb11} we employ the Lorenz gauge to
perform a detailed study of the effects of magnetic fields
in the evolution of binary BHNSs.  

Despite the excellent behavior observed for BHNS simulations when using the
Lorenz gauge, this gauge choice may not be the optimal one for all situations.  
For example, we find that the algebraic gauge yields a better result in the 
magnetized Bondi accretion problem with radial magnetic fields. This is likely 
due to the fact that the system is stationary and the evolution in the 
algebraic gauge rigorously preserves the stationary property: 
$\partial_t A_i = \epsilon_{ijk} v^j B^k=0$ for radial $v^i$ and $B^i$. 
It is therefore useful to explore different EM gauge conditions for 
different problems. Alternatively, one may consider evolving $B^i$ directly 
and construct divergence-free interpolation schemes between refinement 
levels~\cite{Balsara01,Balsara09,McNally11}, 
although implementation of these schemes may be less straightforward. 

\acknowledgments

This paper was supported in part by NSF Grants 
PHY06-50377 and PHY09-63136 as well as NASA Grants NNX07AG96G and NNX10AI73G to the 
University of Illinois at Urbana-Champaign. Z. Etienne gratefully acknowledges 
support from NSF Astronomy and Astrophysics Postodoctoral Fellowship AST-1002667.

\bibliography{paper}

\end{document}